\documentclass[11pt,letterpaper]{article}
\usepackage[T1]{fontenc}
\usepackage[utf8]{inputenc}
\usepackage{times}
\usepackage[margin=1in]{geometry}
\usepackage{microtype}
\usepackage{booktabs}
\usepackage{array}
\usepackage{tabularx}
\usepackage[hyphens]{url}
\usepackage[authoryear,round]{natbib}
\usepackage{hyperref}
\hypersetup{
  colorlinks=true,
  linkcolor=black,
  citecolor=black,
  urlcolor=blue,
  pdfauthor={Kai Yao},
  pdftitle={When AI Does the Work, What Is Learning For? Post-Instrumental Learning and the Risk of Capacity Dissolution},
  pdfsubject={Preprint accepted at AIES 2026},
  pdfkeywords={post-instrumental learning, capacity dissolution, generative AI, assessment, accountable delegation}
}
\newcolumntype{Y}{>{\raggedright\arraybackslash}X}

\title{When AI Does the Work, What Is Learning For?\\
Post-Instrumental Learning and the Risk of Capacity Dissolution}
\author{Kai Yao\\
\small School of Informatics, University of Edinburgh\\
\small Edinburgh, United Kingdom\\
\small \href{mailto:kai.yao@ed.ac.uk}{kai.yao@ed.ac.uk}}
\date{}

\begin{document}
\maketitle

\begin{abstract}
As AI systems become capable of producing the essays, code, reports, summaries, plans, and decisions through which institutions usually recognize competence, a familiar question becomes harder to answer: what is learning for? Existing AI ethics rightly emphasizes present failures--bias, opacity, hallucination, labor extraction, privacy risk, and weak accountability. But if the case for learning rests only on those failures, then each technical improvement appears to weaken it. This article develops a different answer. Using the idealization of AI that executes specified tasks flawlessly while lacking authority over purposes, legitimacy, and responsibility, we argue for \emph{post-instrumental learning}: learning that preserves the capacities people and institutions need when many useful outputs can be delegated. We analyze five such capacities--end-setting, reason-giving, contestability, \mbox{refusal/revision}, and participation--and name their erosion \emph{capacity dissolution}. The central case is assessment under generative AI. When a polished artifact no longer reliably evidences understanding, institutions must assess the learner's accountable relation to AI-mediated work rather than the artifact alone. The takeaway is practical: AI governance should evaluate not only whether systems perform well, but also whether their deployment leaves people able to understand, challenge, revise, and share responsibility for the practices those systems mediate.
\end{abstract}

\section{Introduction}

A student submits a fluent essay and cannot explain its central claim. A manager approves an elegant market analysis without knowing which assumptions make it plausible. A public agency sends an applicant a clear, consistent, and formally correct decision, yet no one the applicant can reach can explain why the rule was framed that way. In each case, the output looks successful. The trouble lies elsewhere: the relation between the person, the institution, and the reasons behind the work has become thin.

This is the anxiety behind many current debates about AI and learning. Generative systems can draft, summarize, translate, code, tutor, classify, plan, and monitor. They can also do these things badly. Current AI systems can be biased, opaque, extractive, environmentally costly, institutionally difficult to contest, and socially harmful. AI ethics has shown how large-scale systems can reproduce inequality, hide responsibility, depend on invisible labor and resources, and deepen relational injustice \citep{bender2021stochastic,crawford2021atlas,ananny2018seeing,burrell2016machine,mittelstadt2016ethics,benjamin2019race,eubanks2018automating,birhane2021algorithmic}. Those criticisms remain indispensable.

But they do not fully answer the educational question. If learning matters only because today's systems hallucinate, discriminate, or fail, then a better system would make learning less important. That conclusion is too quick. It treats learning as preparation for producing outputs, and it treats AI improvement as a reason to produce those outputs with less human effort. Much of that will be true. Some exercises will disappear; some skills will narrow; some forms of practice will migrate to specialists or tools. The harder question is what remains when those changes are granted.

The intuition of this paper is simple: when AI can do the visible work, learning matters because it keeps people in contact with the purposes, reasons, alternatives, and responsibilities behind the work. A student need not write every sentence unaided to learn how to order evidence. A doctor need not calculate every risk score by hand to exercise clinical judgment. A citizen need not master every administrative rule to challenge a public decision. Yet in each case, delegated work remains accountable only if someone can explain what was delegated, why the result should count, and how the practice can be challenged or revised.

We call this form of formation \emph{post-instrumental learning}. The term does not mean that learning has no instrumental value. It marks a shift in emphasis. Once AI can produce many useful artifacts, the central question is no longer whether people must always produce those artifacts without help. It is whether people and institutions retain the capacities to set ends, ask for reasons, contest decisions, refuse harmful dependence, revise practices, and participate in the forms of life AI helps mediate.

The corresponding risk is \emph{capacity dissolution}. This is not ordinary deskilling. A person may stop doing a routine calculation without losing meaningful agency. Capacity dissolution occurs when the practices that make delegated work answerable begin to disappear: the capacity to recognize what has been delegated, judge whether the delegation is legitimate, and intervene when the system acts in one's name or on one's behalf. It can happen while surface performance improves. Cleaner essays, faster classifications, smoother services, and better-formatted reports may conceal a weaker ability to understand and challenge them.

The paper develops the argument through the deliberately generous case of technically successful AI. By ``perfect AI,'' we mean perfect task execution once goals, constraints, and roles have been specified. We do not grant the system authority to decide which purposes should govern, whose interests should count, what forms of dependence are acceptable, or who must answer for the result. That boundary is the point. Even flawless execution does not settle the prior and surrounding questions through which a task becomes legitimate.

The contribution is threefold. First, the paper distinguishes task obsolescence, skill obsolescence, and the stronger claim that formative practices have become obsolete. Second, it gives a positive account of post-instrumental learning through five capacities: end-setting, reason-giving, contestability, \mbox{refusal/revision}, and participation. Third, it applies the account to assessment under generative AI and to AI governance more generally. The practical takeaway is that institutions should ask not only whether AI-mediated work is accurate, safe, or efficient, but also who remains able to understand, contest, and reshape the practice after delegation becomes normal.

\section{Method, Scope, and the Idealization}

The argument is a normative conceptual analysis. It is not a prediction that perfect AI is near, an empirical measurement of learning loss, or a design proposal for one educational technology. The idealization is used as a stress test. It asks whether the case for learning survives after many familiar AI failures have been removed.

By ``perfect AI,'' we mean perfect \emph{task execution after goals have been specified}. Given a goal, constraints, and an institutional role, the system chooses effective means and avoids factual, privacy, safety, and bias failures. It can model consequences, compare options, and explain how an output follows from the assigned objective. What it does not have, by stipulation, is rightful authority to decide which purposes should govern, whose interests should count, what dependencies are acceptable, or who must answer for the result.

The distinction is not a technical footnote; it is the hinge of the argument. A system that chose legitimate ends, distributed responsibility, secured consent, and supplied meaningful recourse would no longer be merely executing tasks. It would occupy the space of moral and political judgment. We therefore grant flawless execution and ask what delegation still requires once performance is no longer the easy objection.

Automation research gives a practical reason for drawing the boundary this way. High system performance does not eliminate residual human responsibilities such as oversight, reliance calibration, intervention, and responsibility allocation \citep{bainbridge1983ironies,parasuraman1997humans,parasuraman2000model,lee2004trust}. Delegation to autonomous AI normally occurs after a problem has already been framed \citep{candrian2022rise}. Institutions translate purposes into rubrics, eligibility rules, prompts, labels, performance indicators, and optimization targets. Systems then operate within those translations. People and procedures still carry problem formulation, authorization, review, appeal, and recourse \citep{passi2019problem,binns2022human,almada2019human}.

Two cautions follow. First, the argument is not a defense of human involvement for its own sake. Human judgment can be biased, corrupt, fatigued, uninformed, or exclusionary. AI can expand agency when it removes barriers, supplies patient tutoring, translates across languages, supports disabled users, or makes expertise more available. Second, learning here is broader than schooling. It includes professional formation, civic practice, workplace apprenticeship, public deliberation, and the institutional routines through which people learn to ask for reasons and challenge decisions.

The principal worked case is formal educational assessment, but this does not define the paper's account of learning or knowledge. Essays, examinations, code, and related artifacts matter because institutions use them as evidence of capacity; the argument does not assume that such artifacts exhaust how learning occurs or can be demonstrated.

The standard is therefore institutional as well as individual. No society can require every person to master every technical operation on which they depend. Modern agency already relies on specialization. Delegation remains accountable, however, only if enough capacity persists across affected people, intermediaries, and public or professional communities. A community may preserve a few experts and still leave most affected people with no usable route to reasons or remedies. It may train users to operate an interface while leaving them unable to judge the goals that interface serves. The relevant question is not whether everyone understands everything. It is whether AI-mediated work can still be connected back to purposes, reasons, alternatives, and repair.

\section{From Obsolescence to Capacity Dissolution}

A great deal turns on a distinction that is often blurred. AI may make a task unnecessary for most people. It may make a technique less worth teaching. Or it may make the learning process itself appear dispensable. These claims sound similar in policy debate, but they point to different consequences.

Task obsolescence is familiar. Most people no longer grind grain, navigate by stars, or calculate with logarithm tables. Skill obsolescence is also real, though rarely complete: old skills may narrow, migrate to specialists, or survive as diagnostic knowledge. A physician may not perform every calculation by hand, but still needs enough understanding to notice when a number is clinically implausible. A lawyer may use search and drafting tools, but still needs enough legal judgment to know which issue is doing the work. The strongest claim, formative obsolescence, is different. It says that the practice through which a capacity was cultivated no longer matters because the output can be obtained without it.

This is where the debate over AI and learning often moves too fast. Because AI can draft essays, solve problem sets, or produce code, students may appear not to need writing, mathematics, or programming in the old way. Sometimes that is right. A particular exercise may be obsolete. But it does not follow that the underlying capacity is obsolete. The better question is what the exercise was doing. Was it merely a route to a product, or was it a route into judgment--learning what counts as evidence, how reasons hang together, where uncertainty matters, and when a result should be resisted?

By \emph{capacity dissolution}, we mean a breakdown in the relation among an output, the person or institution that claims or relies on it, and the practice that makes that claim answerable. The artifact, decision, or service remains; the capacity to explain, contest, or repair it fades.

Capacity dissolution is related to, but not reducible to, three neighboring concerns. Deskilling concerns diminished ability to perform a task independently \citep{vallor2015moral}; automation complacency concerns poorly calibrated reliance and monitoring \citep{parasuraman1997humans,lee2004trust}; and diminished contestability concerns whether affected people have usable means to challenge a decision \citep{alfrink2023contestable,lyons2021conceptualising}. Capacity dissolution may coexist with any of these failures, but its object is the institutional relation that keeps delegated work answerable. It can arise while specialists retain the relevant skill, system performance remains high, and a formal appeal route exists, if people no longer develop the judgment needed to set ends, use reasons, or make challenge consequential. Its distinctive warning sign is successful output production alongside the erosion of the learning practices that sustain accountability.

The mechanism is easy to miss because the first signs often look like improvement. Schools receive cleaner essays. Hospitals receive more complete discharge summaries. Public agencies make faster classifications. Workplaces produce better presentations, reports, and plans. Nothing in the immediate output announces that the surrounding practice has thinned out. The loss appears later, when a student cannot defend a claim, a clinician cannot reconstruct a recommendation, a caseworker cannot explain an eligibility decision, or a manager cannot tell whether a generated plan serves the right goal.

The issue is therefore evidentiary as well as practical. Many artifacts have long served as evidence of capacity. An essay stood for understanding; a proof for mathematical command; a case note for professional judgment; a policy memo for grasp of alternatives. Generative AI loosens this evidentiary relation. The institution can still receive the artifact, but the artifact no longer shows what it used to show. If the institution continues to treat the product as proof of competence, it mistakes output preservation for capacity preservation.

There is also a political dimension. Capacity is not a private stock that survives untouched after it has once been acquired. It is often sustained by recurring occasions for use and by institutions that make its use consequential. If acceptable outputs keep arriving, pressure will favor further delegation. The cost of lost capacity usually appears only when someone must explain, repair, contest, or refuse the system. By then the relevant people may no longer have practiced the needed form of judgment, and the institution may have redesigned its workflow around the assumption that such judgment is rarely needed. Capacity loss is thus cumulative: each successful delegation can remove another occasion for learning how to govern delegation.

This is why capacity must be distributed, not merely kept inside an expert enclave. The phrase ``practical right to understand'' is used here as a governance standard, not as a fully specified legal right. People affected by a system need usable routes to reasons, recourse, and revision. They do not need to understand every technical detail, but they do need access to people, records, procedures, and public forums that can connect their problem to an answerable decision. A black-box system can fail this standard; so can a formally transparent system whose explanation is useless to the people who must live with it.

A democratic analogy helps. Citizens delegate lawmaking to representatives, but a democracy in which citizens lose all capacity to understand, criticize, replace, or organize around representatives is democratic only in name. Delegation is compatible with agency because supporting capacities remain: public argument, journalism, education, opposition, courts, associations, and ordinary practices of explanation. The AI analogue is not universal expertise. It is a learning ecology in which some people can build systems, others can interpret them, affected people can challenge them, and institutions can revise them.

Capacity dissolution therefore changes the time horizon of AI governance. Many policies ask about the immediate quality of a decision: Was it accurate? Was it biased? Was private information protected? Those questions remain essential, but they are not enough. Institutions must also ask what repeated reliance teaches people not to do. Does it reduce opportunities to formulate questions, inspect reasons, compare alternatives, handle exceptions, or participate in the practice behind the service? The harms may not show up as a single bad decision. They may appear as a public, profession, or classroom that becomes less able to understand why the decisions are being made at all.

\section{Post-Instrumental Learning}

Post-instrumental learning is the formation of capacities that remain necessary for agency, responsibility, and meaning under extensive AI delegation. It is not a claim that learning has no practical payoff. It is a claim about what the practical payoff becomes when many products of learning can be delegated. The question shifts from ``Can a person produce this output?'' to ``Can the person or institution stand behind, challenge, and reshape the work that produced it?''

Five capacities are central: end-setting, reason-giving, contestability, \mbox{refusal/revision}, and participation. They form a proposed minimum functional set, not a complete theory of education. The claim is normative and functional rather than psychometric: the capacities may overlap, and we do not present them as an empirically validated scale. Their selection follows the functions required across a delegation cycle. End-setting keeps purposes open to judgment before and during delegation. Reason-giving connects an adopted output to considerations others can inspect. Contestability supplies a route to challenge that output or its justification, while refusal/revision makes exit and redesign possible when dependence itself becomes unacceptable. Participation sustains the shared practices in which these capacities are learned and standards are renewed. Omitting one can leave a distinct point at which delegation becomes abdication.

\emph{End-setting}. Delegation begins with the question of what the system is for. End-setting is the capacity to articulate and revise goals rather than accept optimized suggestions as given. It includes the ability to distinguish first-order desires from reflective commitments, to notice when a metric has displaced a purpose, and to ask whether an institution has defined the problem in the right way. A system can optimize a triage rule, a tutoring plan, or a productivity target; it cannot by technical success alone determine whether that target deserves authority.

\emph{Reason-giving}. An output becomes answerable when someone can say why it should be trusted, why an action is justified, and why a trade-off is acceptable. Reason-giving is not the same as technical transparency. It is the ability to connect an output to considerations that the relevant audience can inspect and challenge. A probability score, feature attribution, or generated explanation may help, but only when it enters a practice in which reasons can be tested rather than merely displayed.

\emph{Contestability}. Affected people need more than a formal right to complain. They need the capacity and institutional standing to ask who benefits, who is burdened, what assumptions are embedded, what data are used, and what remedy is available. In domains structured by race, class, disability, gender, and migration status, contestability also requires sensitivity to epistemic injustice: whose testimony is discounted, whose categories are imposed, and whose experience is treated as noise \citep{fricker2007epistemic,benjamin2019race,eubanks2018automating,birhane2021algorithmic}.

\emph{Refusal/revision}. This paired capacity is the ability to say no to automation, exit a workflow, and rebuild a practice when dependence becomes harmful. The two terms belong together because refusal preserves agency only if people or institutions can do something after refusing. A nominal opt-out that leads to delay, worse service, suspicion, or practical exclusion is not meaningful refusal. Revision matters because some dependencies become unacceptable not through one dramatic failure, but through the gradual disappearance of alternatives.

\emph{Participation}. Some practices matter because they form standards, mutual recognition, and practical judgment. These goods become intelligible through shared practice, not through efficient delivery alone. A finished song, diagnosis, essay, judgment, or policy recommendation may be valuable, but the practice around it also teaches people how to listen, object, justify, revise, and take responsibility before others. Participation names this formative and relational dimension of learning.

Table~\ref{tab:pil} translates the five capacities into institutional design questions. The capacities are interdependent but not interchangeable. Goals require reasons; reasons need routes for challenge; refusal requires the ability to revise; participation gives people a setting in which the other capacities can be practiced. The practical takeaway is that institutions should design not only for better outputs, but also for recurring occasions to exercise these capacities.

\begin{table}[t]
\small
\centering
\caption{Post-instrumental capacities and their institutional failure modes.}
\label{tab:pil}
\setlength{\tabcolsep}{4pt}
\begin{tabularx}{\textwidth}{@{}p{.15\textwidth}YY@{}}
\toprule
Capacity & Institutional design question & Failure if capacity disappears \\
\midrule
End-setting & Who can define, question, and revise the goals an AI system serves? & Optimized means drift away from legitimate purposes. \\
Reason-giving & Can affected people obtain reasons they can understand, test, and use? & Explanations become display rather than accountability. \\
Contestability & Is there a real path to challenge assumptions, outputs, and burdens? & Affected people encounter decisions as settled facts. \\
Refusal/revision & Can people refuse, exit, or redesign a workflow without hidden penalty? & Dependence becomes compulsory even when harmful. \\
Participation & Does the system preserve occasions to practice judgment with others? & Outputs remain, but the practice that made them meaningful thins out. \\
\bottomrule
\end{tabularx}
\end{table}

The capacities should be distributed across affected people, institutional intermediaries, and public or professional communities. Affected persons need routes to stakes, reasons, and help. Intermediaries such as teachers, clinicians, caseworkers, lawyers, engineers, and public officials need enough independent judgment to resist becoming rubber-stamps. Communities need shared forums in which standards can be argued over and revised. This distribution is what separates accountable delegation from private dependence on expertise.

This is also why AI literacy is insufficient on its own. Competent users cannot repair systems that give them no role, and appeal processes have little value when people have not learned how to use them. The delegation chain needs users who can ask better questions, intermediaries who can exercise judgment, and institutions that make challenge consequential. Learning is therefore not a personal upgrade added after deployment. It is part of the infrastructure that makes deployment answerable.

The account extends older work on practical judgment, education, and formation. Aristotle treats practical judgment as a cultivated disposition rather than an item of information \citep{aristotle1999nicomachean}. Dewey understands education as the reconstruction of experience through participation in shared problems \citep{dewey1916democracy,dewey1938experience}. Communities of practice show that expertise is learned through situated participation rather than detached rule following \citep{lave1991situated,wenger1998communities}. Work on technological mediation and extended agency likewise shows that tools shape perception, action, and responsibility rather than simply adding capacity from the outside \citep{ihde1990technology,clark1998extended,hayles1999how,vallor2016technology}.

This relational view also places the account alongside postdigital work, which treats educational practice as inseparably digital, non-digital, material, and social, and critical posthuman work, which questions the autonomous human learner as education's default subject \citep{fawns2019postdigital,bayne2017posthuman}. Distributed cognition similarly locates cognitive activity across people, artifacts, and environments \citep{hollan2000distributed}. These are relational starting points for our account, which asks a further governance question: what learned capacities must remain when distributed activity becomes routine AI delegation if the resulting practice is to stay answerable? The perfect-AI idealization likewise uses an imagined future to expose present assumptions rather than predict technical development, a role shared with speculative work in higher education \citep{bayne2024speculative}.

It also connects AI literacy and AI education to governance debates over meaningful human control, contestability, reviewability, and recourse \citep{long2020ai,ng2021conceptualizing,molenaar2025education,wu2025epistemic,markauskaite2022rethinking,santoni2018meaningful,alfrink2023contestable,cobbe2021reviewable,karimi2021recourse,venkatasubramanian2020recourse}. The contribution here is to put those debates under a learning question: what capacities must a person, profession, institution, or public continue to develop if delegated work is to remain answerable?

\section{Four Limits Perfect AI Execution Cannot Remove}

Perfect execution solves problems of means. It does not settle what counts as a worthy end, whether a decision has standing, whether integrity has been preserved, or whether participation survives. Each of the following limits answers a tempting claim: if AI is accurate, safe, detectable, or experientially rich enough, perhaps learning can retreat. The limits show why that retreat would be premature. They do not show that AI should be used less. They show where AI use must be accompanied by practices that preserve judgment.

\subsection{Accuracy and End-Setting}

Accuracy is an epistemic good. It concerns how well a system represents facts, predicts consequences, or reasons from premises. But an accurate map does not choose the destination. Many AI-mediated decisions are not settled by facts alone. A city may ask whether to optimize for traffic flow, pedestrian safety, emissions, housing access, revenue, or neighborhood continuity. A school may ask whether to maximize test scores, curiosity, inclusion, disciplinary mastery, or civic preparation. A workplace may ask whether productivity means output per worker, better service, less burnout, or more equitable opportunity.

AI alignment research makes this point visible. Alignment is a normative question about which instructions, preferences, values, and procedures should count, as well as a technical question about getting systems to follow instructions \citep{gabriel2020artificial,russell2019human}. Even a perfectly accurate model can carry forward a bad problem formulation. It can show which means best serve a target, but it cannot make the target legitimate by serving it well.

End-setting is itself learned. People learn to distinguish desire from reason, impulse from commitment, proxy from purpose, and efficiency from value. They also learn to see when the available options have been framed too narrowly. If AI systems increasingly present goal formulations as natural, ranked, or ready for execution, people may lose practice in the prior activity of asking what should be pursued. The risk is not that the system gives a wrong answer. The risk is that people forget that the question could have been otherwise.

\subsection{Safety and Legitimacy}

Safety means that a system avoids certain harms. Legitimacy asks whether a decision has standing: whether it is authorized by appropriate procedures, responsive to those affected, and answerable to public or professional norms. A system can be safe in the narrow sense and still lack legitimacy. It may produce correct benefit classifications while affected people have no meaningful way to challenge the rule. It may allocate resources without discrimination while the affected community never had a voice in the goal. It may recommend clinical care that is statistically sound while leaving the patient unable to understand or contest the trade-off.

This distinction is central to algorithmic accountability. Transparency, explanation, human review, and recourse are not valuable merely because systems sometimes err. They are valuable because they help locate authority, expose assumptions, and make decisions revisable \citep{ananny2018seeing,binns2022human,wachter2018counterfactual,barocas2020hidden,karimi2022survey}. A world of more reliable AI would still need these practices, because legitimacy is not reducible to error correction.

Learning matters here because legitimacy is practiced. People learn how to give reasons, hear objections, recognize standing, and revise decisions under pressure from others. Institutions also learn: they develop routines for appeals, exceptions, consultation, and public justification. If AI removes too many occasions for these practices, the institution may retain safe outputs while losing the habits by which its authority is kept answerable.

\subsection{Detection and Integrity}

A common institutional response to generative AI is detection: determine whether an artifact was produced by a human or a machine, then enforce the rule. Detection may sometimes be necessary. It is not the deepest issue. When AI can produce fluent essays, code, reports, designs, and analyses, the ethical problem is not only authorship. It is whether the artifact still evidences the capacity an institution claims to certify.

Integrity is therefore not just the absence of cheating. It is a relation among learner, task, tool, and reason. A student who uses AI to generate a counterargument, then revises a claim in light of that challenge, may be exercising more judgment than a student who writes a shallow paragraph unaided. Conversely, a student who submits a polished AI-generated essay without understanding it has not demonstrated the capacity the artifact appears to display. The relevant question is not simply ``Was AI used?'' but ``What did the learner remain responsible for?''

This reframing matters because detection-centered regimes can weaken the very capacities they seek to protect. They may encourage adversarial compliance, surveillance, and mechanical prompt records while leaving students unclear about why a task matters. They may also burden students unevenly, especially those who rely on assistive technologies, language support, or legitimate scaffolding. Assessment should protect integrity by making the target capacity visible, not by treating every polished artifact as a suspected violation.

\subsection{Experience and Participation}

Some goods are lost when a process disappears, even if the result can be obtained. The claim is not romantic nostalgia for difficulty. It is that practices often contain forms of perception, timing, mutual adjustment, and responsibility that are learned only by taking part. A case conference teaches junior clinicians how uncertainty is handled. A seminar teaches students how reasons change in response to objection. A public hearing teaches participants what it means to make a claim before others who can answer back.

Generative AI can support these practices. It can simulate objections, provide examples, translate, summarize, and make participation easier for people who were previously excluded. The risk arises when simulation replaces the setting in which standards are shared and contested. A student who receives a perfect explanation may still lack the experience of making a claim and having it challenged. A public agency that produces polished consultations may still fail to create a forum in which affected people can reshape the issue.

Participation therefore marks a limit of output-based thinking. A practice can be valuable because of what it produces and because of what it does to the people who participate in it. The first value is easier to automate. The second is easier to overlook. Post-instrumental learning asks institutions to preserve or redesign the participatory settings through which people become capable of judgment, not to preserve every inherited inconvenience.

These four limits show why the idealization sharpens rather than weakens the case for learning. Accuracy, safety, detection, and efficiency are important. But none of them by itself secures the capacities that make delegated work answerable. Once AI performs well, those capacities must be designed for deliberately; they will not be preserved by the output alone.

\section{Assessment After Generative AI}

Assessment is the clearest case of capacity dissolution because schools and universities routinely use artifacts as evidence of learning. An essay, code submission, lab report, proof, or design portfolio is not only a product. It is a proxy for a learner's understanding, judgment, and ability to work within a discipline. Generative AI weakens that proxy. The artifact can remain impressive while the evidentiary relation changes.

\paragraph{Validity before enforcement.}

A policy that begins with cheating starts too late. The first question is validity: what capacity is this assessment meant to evidence, and does the artifact still provide credible evidence of it? Assessment scholarship has increasingly emphasized this point in response to generative AI \citep{dawson2024validity,bearman2024developing,lodge2023assessment,weng2024assessment}. A take-home essay may still assess research judgment if students must frame a question, justify source choices, explain revisions, and defend claims. The same essay may no longer assess those capacities if the only evaluated object is a polished final text that could have been generated without understanding.

This validity question is not a reason to abandon product assessment. Products matter. A nurse's note, legal brief, policy memo, software patch, or research article must still be usable. But if an institution claims to certify a person's capacity, it must test the relation between the person and the product. The evidence may need to move partly into oral defense, process commentary, source comparison, in-class transfer, version histories, or tasks in which students respond to new constraints. These are not anti-AI measures. They are ways of restoring the evidentiary link between artifact and capacity.

\paragraph{Three assessment models.}

Product-only assessment asks whether the submitted artifact meets a standard. This model fits situations in which the institution genuinely cares only about the product or in which AI use is already part of the professional performance being assessed. Its weakness is that it can certify the appearance of capacity without capacity. A student may submit a correct program, elegant essay, or polished analysis and still be unable to explain the problem, interpret the evidence, or adapt the solution.

Authorship enforcement asks whether the artifact was produced by the student under specified conditions. It protects some forms of practice, especially where unaided fluency, memory, or procedural command remain legitimate goals. Its weakness is that it can make detectable human production the point of the assessment. It also tends to invite surveillance and uneven suspicion, while neglecting cases in which AI use genuinely supports learning or accessibility \citep{liang2023detectors,perkins2024aias,miao2023guidance}.

Post-instrumental assessment asks whether the learner can stand in an accountable relation to AI-mediated work. The relevant evidence is not merely the final artifact and not merely the absence of AI. It is the learner's ability to frame the task, make judgments about the tool's contribution, evaluate evidence, revise claims, answer objections, and transfer understanding to a new case. This model is stricter than permissive product assessment, because a polished artifact is not enough. It is also less brittle than blanket prohibition, because it can distinguish replacement from scaffolding.

\paragraph{Designing accountable use.}

A practical design sequence follows. First, name the target capacity. Is the task assessing source evaluation, conceptual explanation, methodological choice, mathematical fluency, design judgment, professional communication, or civic argument? Second, decide which forms of AI use would scaffold that capacity and which would replace it. Third, require evidence at the point where judgment matters: a brief justification of source selection, a defense of a contested premise, a comparison between generated and primary-source claims, or an explanation of why an AI suggestion was rejected.

The documentation should be lean and meaningful. Asking students to paste every prompt can generate a large record that no one reads and that says little about judgment. Better evidence focuses on responsibility: where the learner changed course, corrected a model, revised a claim, checked a source, or chose among alternatives. A short process note, targeted annotation, viva, or in-class transfer task can do more than exhaustive prompt logs because it asks the student to leave the prepared artifact and reason under a new demand.

Accountable-use design also requires equity checks. A policy that treats all AI use as suspicious may punish disabled students, multilingual writers, and students who use legitimate support. A policy that treats all AI use as harmless may benefit students who already know how to direct tools strategically while leaving others with shallow completion. The capacity question makes the equity issue more precise: which students are being given opportunities to develop judgment, and which are being given only finished products, surveillance, or prohibition?

These design principles do not require every assessment to become elaborate. Some tasks should remain AI-free because fluency, memory, or embodied practice is the target. Some tasks should be AI-integrated because professional work already requires tool-mediated judgment. Some tasks should be redesigned or removed because they no longer provide useful evidence. The point is not to choose a universal rule. It is to align the rule with the capacity the assessment claims to certify.

Assessment makes the paper's general problem concrete. Once AI can produce the artifact, an institution must decide whether it is evaluating the artifact alone, enforcing an authorship rule, or certifying the learner's accountable relation to the work. The third option is post-instrumental. It asks whether delegation has preserved the learner's capacity to understand, defend, revise, and take responsibility for what is submitted.

\section{Delegation Without Abdication}

Education makes the evidentiary problem unusually visible, but the same structure appears wherever institutions delegate consequential work to AI. Human life already depends on delegation: we rely on pilots, translators, judges, farmers, engineers, nurses, teachers, public officials, and infrastructures we barely understand. The ethical question is not whether delegation is acceptable. It is how institutions can delegate to AI without surrendering the capacities that make delegation accountable.

\paragraph{Conditions for accountable delegation.}

AI delegation expands agency when people can still shape goals, connect outputs to usable reasons, object and obtain remedies, participate in relevant practices, and fall back when dependence becomes dangerous. These conditions need not be held by one person, and their stringency should scale with stakes. They belong in routines, review processes, training, and procurement decisions as much as in interface design.

\paragraph{When friction is formative.}

The distinction between exclusionary and formative friction is crucial. Exclusionary friction includes needless paperwork, inaccessible interfaces, opaque appeals, and mandatory effort that burdens people who already have less time, money, or institutional power. Formative friction asks users to state a goal before optimizing, compare alternatives before accepting a recommendation, confront a counterargument before submitting a claim, or explain a choice before delegating a high-stakes action.

Formative friction should be proportionate and accessible. The burden should match the stakes, and it must not exclude disabled users, multilingual users, overburdened workers, or people with limited resources. Users should also be able to challenge or adapt friction when it is misplaced. These standards keep necessary difficulty from becoming a polite name for exclusion.

Post-instrumental learning does not require pure human control. Agency is often relational, exercised through people, tools, institutions, and environments \citep{clark1998extended,hayles1999how,haraway1991simians}. Responsibility still needs trained bearers: people and institutions able to answer for what delegated systems do. Relational agency complicates responsibility; it does not make responsibility disappear.

\paragraph{Maintaining capacity.}

Delegation requires maintenance. A hospital, school, court, newsroom, or public agency can adopt an AI system and still preserve human capacity, but only if it treats capacity as part of the system being maintained. Newcomers need hard cases. Experienced workers need time to review borderline outputs instead of becoming passive signers. Affected people need reasons that can change outcomes. Institutions should record where human judgment corrected the system, where system use revealed a routine that should change, and where dependence became too deep to explain in ordinary terms.

This point matters because abdication often appears as a sequence of reasonable local decisions. A teacher uses a model to draft feedback because the class is too large. A manager accepts generated summaries because meetings are too many. A public office automates triage because staff are overwhelmed. Each choice may be defensible on its own. The pattern becomes troubling when no one asks which capacities will still be practiced, by whom, and under what conditions. Post-instrumental learning makes that question visible before dependence becomes naturalized.

\section{Implications for AI Ethics and Governance}

Accuracy, robustness, privacy, fairness, and explainability remain essential. The argument adds another governance object: the learning ecology around a system. Institutions should ask not only whether a deployment produces acceptable outputs, but also how it changes the distribution of capacity among designers, operators, intermediaries, and affected people. The question is practical: after the workflow settles, who can still understand, challenge, and revise what the system is doing?

\paragraph{Effects on agency.}

Capacity effects are often indirect. A system may not coerce anyone, misclassify anyone, or expose private data, yet still reduce the opportunities through which people learn how a practice works. It may make the novice less able to develop judgment, the professional less able to notice misfit, or the public less able to organize around a contested institutional choice. Such losses are easy to miss because they appear later, when explanations become harder to use, appeals rarely change outcomes, fallback practices fade, and ordinary challenge starts to seem unrealistic.

Institutional review should begin with the capacities in Table~\ref{tab:pil}: which capacity is at stake, how AI changes opportunities to exercise it, whether any formative friction is proportionate and accessible, and who gains or loses the chance to learn. The review should ask whether a system performs well and whether it leaves people more able to understand, question, refuse, revise, and share responsibility for AI-mediated work.

\paragraph{Distributional audits.}

A distributional audit asks who becomes more capable, and who becomes less capable, through a system. Here, ``audit'' means a structured institutional check, not only a statistical compliance audit. A lawyer may draft faster while retaining enough expertise to notice errors; a client may receive smoother service while understanding less about the legal choices being made. A teacher may generate examples while retaining curricular judgment; a student may use the same system to bypass the struggle through which that judgment would form. The same interface can scaffold one role and reduce another role to output receipt.

The audit should therefore track opportunities to learn as well as access to a tool. Access may mean receiving a finished answer. It may also mean learning how the answer was framed, what it excluded, and how it might be challenged. These are different goods. A public benefits chatbot may speed service while leaving applicants unable to understand a classification. A workplace assistant may ease reporting while reserving strategic understanding for template setters and exception reviewers. An educational platform may personalize practice while steering disadvantaged learners toward narrow completion and advantaged learners toward open-ended coaching. The audit question is simple: who gets to become more capable through the system, and who is kept at the level of receiving outputs?

\paragraph{Assessment policy.}

The assessment case can be restated as a governance procedure: assess the learner's accountable relation to AI-mediated work. Policy should begin with the capacity the artifact is supposed to evidence. It should then specify how students can demonstrate that capacity, what AI uses scaffold or replace it, what evidence is accessible, and whether the policy gives some students chances to learn while giving others only suspicion, surveillance, or finished products.

This moves assessment away from the binary of AI allowed or forbidden. The better question is: what relation between learner, tool, artifact, and reason is this assessment meant to certify? An institution that cannot answer that question has not solved the problem by banning AI, requiring disclosure, or approving unrestricted use. It has simply moved uncertainty into enforcement.

\paragraph{A practical right to understand.}

For institutions, the practical right to understand has a concrete form. They should identify where people need reasons they can use, where they need a human intermediary with independent judgment, and where they need a fallback path that does not punish refusal. They should also ask whether their AI policy changes who gets to learn the practice. If privileged users learn to direct, audit, and contest AI while marginalized users receive automated completion or automated suspicion, the institution has changed the distribution of expertise while calling the result access.

This practical right should not be confused with a demand that everyone master every technical detail. A patient need not train a diagnostic model to ask why a recommendation fits their case; a tenant need not understand an allocation algorithm to challenge a housing decision; a student need not reproduce every step of a model's generation to defend a claim made in a paper. Each person needs a way to move from a problem they experience to reasons they can understand and, where appropriate, to a process that can revise the decision.

That path depends on trained intermediaries, accessible records, meaningful appeal, and public settings where reasons can be tested. It fails when explanation exists only for specialists, when refusal carries hidden penalties, or when participation is reduced to accepting a finished service. Expertise has to remain distributed enough to be answerable: some people must understand how systems are built, others how they are used, how they fail, and how affected communities experience them. Governance fails when these forms of knowledge are separated so sharply that no forum can put them into conversation.

\paragraph{Deployment review.}

The central review question before deployment is simple: what will people stop learning if this system works? The answer may be harmless. People can stop learning many obsolete routines without loss. But if they stop learning how to set goals, give reasons, contest decisions, refuse dependence, or participate in a valued practice, the institution should redesign the workflow. It might preserve a human explanation step, require periodic manual practice, rotate staff through review roles, support public audit, or create spaces for baseline formation without AI assistance. Such measures make delegation more durable by preserving the capacities on which the tool depends.

The same question is useful after deployment. Institutions should watch whether the learning ecology around the system is changing: whether junior workers still see hard cases, teachers can still see students reason beyond generated artifacts, affected publics receive usable explanations, appeals produce revisions, and fallback practices remain real. These are ordinary governance questions once capacity is treated as part of the system being governed.

A lightweight capacity impact statement can make this review concrete. It records five judgments. The first three identify (1) the changed practice and responsible roles; (2) the capacity the prior practice formed or evidenced; and (3) who needs that capacity or relies on its certification. The final two specify (4) how the new workflow preserves an opportunity to exercise it and what evidence makes that opportunity credible, and (5) an observable trigger for review or repair. This is a deliberative record, not a validated scale or universal score. Its purpose is to make the institution's reasons inspectable and revisable.

Consider an AI-supported research essay designed to assess reason-giving through source evaluation. AI can search, summarize, and draft, so the final essay no longer shows whether the student judged a source's authority, relevance, and limitations. Students need that judgment, and instructors need credible evidence before certifying it. The redesigned task can require a brief justification of key source choices, comparison of a generated claim with its primary source, and a response to an unseen objection. If essays improve but students repeatedly cannot defend source selection or revise their argument when evidence changes, the review trigger has been met. The institution should then revise the task or restore guided practice in source evaluation rather than infer capacity from the product.

Institutions need not create a separate compliance process for every deployment. The statement can be incorporated into course approval, assessment moderation, procurement, or periodic review, with its depth and review interval scaled to the stakes and reversibility of the workflow. A low-stakes drafting aid may require only a brief rationale; consequential certification requires direct evidence and continued monitoring. The point is to distinguish parts of the old practice that were dead weight from parts that carried judgment, then preserve a usable route to learning, challenge, and repair.

The time horizon of AI ethics therefore has to widen. Much ethical analysis asks whether a system respects rights, distributes benefits fairly, or avoids harm at a point in time. Post-instrumental learning adds a question about what kinds of agents and publics the system helps produce over time. A tool may be fair in its current outputs, explainable in principle, or accessible as a service while still making future contestation harder. The response is not to reject automation, but to treat capacity maintenance as one of the outcomes of design.

Some apparently inefficient practices should therefore be preserved or redesigned rather than simply removed. A classroom discussion, peer review meeting, case conference, or public hearing may look slow compared with AI-assisted synthesis. Its value lies partly in the output, but also in the formation of people who can listen, object, revise, and take responsibility before others. Empty friction blocks access without improving agency. Formative friction gives people a real opportunity to exercise a capacity that matters.

\section{Objections and Clarifications}

Several objections help sharpen the account. The first is that capacity maintenance may sound like a demand for universal expertise or endless human labor. It is not. The claim is not that every student, patient, client, worker, or citizen must be able to reproduce an AI-mediated output unaided, or that institutions should preserve manual work for its own sake. That would misunderstand both modern specialization and the role of institutions. The claim is that delegation needs a usable chain of understanding. Someone must be able to connect the affected person's question to reasons, exceptions, and remedies; someone must have enough independence to challenge the system; and the affected person must have standing to activate that chain. Capacity dissolution occurs when each link assumes that another link still understands the work, while the person at the end encounters only the finished result.

The second objection is that preserving capacity may become a conservative defense of old routines. This is a real danger. Many inherited practices were exclusionary, inefficient, or poorly connected to the capacities they claimed to form. Requiring students to perform every step unaided can punish disability, language difference, poverty, and lack of prior preparation. Requiring workers to preserve manual routines can waste time and reinforce hierarchy. Post-instrumental learning does not treat difficulty as good in itself. It asks whether a practice gives people a real opportunity to exercise a capacity that matters. When the answer is no, AI should remove the burden. When the answer is yes, institutions should redesign the practice so that the formative opportunity remains accessible.

The third objection is that AI itself may support the capacities described here. That is also true. AI can make reasons more available, generate counterarguments, simulate consequences, translate expert language, and help learners practice beyond the limits of classroom time. The distinction is not between human and machine contribution. It is between scaffolding and replacement. A scaffold helps a person enter a practice, see its standards, and exercise judgment with support. Replacement supplies the outcome while removing the occasions on which judgment would be formed. The same tool can do either, depending on the surrounding task, incentive, and institutional response.

This distinction explains why the account is not anti-automation. A well-designed AI tutor might strengthen post-instrumental learning by asking a learner to explain a step before receiving feedback, by offering contrasting arguments, or by adapting examples so that a disabled or overburdened learner can participate more fully. A poorly designed system might weaken the same capacity by completing the assignment, hiding the reasoning, and training the learner to treat acceptance as success. The ethical difference is not located in the model alone. It is located in the relation among tool, task, learner, and institution.

A fourth clarification concerns thresholds. Not every delegated routine needs preservation. People can stop learning obsolete techniques without serious loss. The capacity question becomes urgent when three conditions coincide: the system mediates a consequential practice; the output is treated as evidence of human or institutional judgment; and affected people need a meaningful route to reasons, refusal, or revision. Under those conditions, capacity maintenance is not optional. It is part of what makes the practice legitimate.

These thresholds are deliberately practical. They allow low-stakes automation to proceed without moral drama while identifying domains where smooth output can hide real abdication. Autocomplete in a casual message raises little concern. Automated drafting of a legal advice letter, a benefit denial, a clinical discharge note, or a student research essay raises more. In each case, the question is not whether AI helped. It is whether the responsible person or institution can still explain the relevant choice, answer a challenge, and repair the practice if the delegation proves harmful.

The account also has a distributional edge. Capacity is not merely lost or preserved; it is shifted. AI may increase the capacity of those who set prompts, design templates, approve exceptions, or interpret system outputs, while reducing the capacity of those who receive completed services. This shift can be desirable when it removes pointless burden and expands access. It is troubling when already powerful actors become better at directing systems while less powerful actors become less able to understand or contest what is done to them. Capacity maintenance is therefore a question of justice as well as pedagogy.

These clarifications return the argument to its central claim. The danger is not that AI will make people use tools. Human agency has always been tool-mediated and socially distributed. The danger is that institutions will confuse successful completion with accountable practice. Post-instrumental learning names the work required to avoid that confusion: preserving enough learned contact with goals, reasons, alternatives, and repair that AI-mediated work remains answerable to those it affects.

\section{Conclusion}

We began with a deliberately generous assumption: AI works. It does not hallucinate, discriminate, or fail in the familiar ways. Even then, learning remains necessary because delegation still needs people and institutions able to name goals, test reasons, challenge errors, refuse harmful dependence, and participate in practices that can be rebuilt. These capacities are not decorative additions to performance. They are the conditions under which performance remains answerable.

This answer does not preserve old exercises unchanged. Some tasks will disappear, and some should disappear. The point is to identify which practices still do formative work after their immediate products can be generated. A writing assignment may matter because it trains students to order reasons; a case conference because it lets junior clinicians see how experienced judgment handles uncertainty; a public hearing because it gives affected people a setting in which reasons can be heard and revised. In each case, the output is only part of the point.

The assessment case shows the issue in miniature. Once generative AI can produce polished essays, code, reports, designs, or analyses, institutions must ask what the assessment certifies: a competent-looking product, unaided production, or the learner's accountable relation to AI-mediated work. Product-only assessment treats the artifact as the achievement. Authorship enforcement treats detectable human production as the achievement. Post-instrumental assessment treats the learner's ability to explain, defend, revise, and take responsibility as the achievement.

The broader ethical problem is abdication through capacity dissolution, and abdication is often quiet. It may look like convenience, personalization, productivity, or expert service. It becomes visible when people need to ask why a goal was chosen, why a reason should be accepted, why a decision should bind them, or how a practice can be rebuilt. A society can keep the outputs and lose the capacities that make the outputs accountable.

Addressing this form of abdication requires neither human exceptionalism nor technological optimism. The framework can support systems that expand participation while still defending difficult practices that train judgment. Empirical research can ask how AI-mediated workflows change novice formation, appeal behavior, professional judgment, and public understanding. Design research can ask how interfaces invite goal-setting, comparison, challenge, and learning from exceptions. Governance research can ask how capacity impact statements, assessment policies, and audit routines can be made concrete without becoming empty compliance exercises.

The paper's most compact takeaway is therefore a change in the default question. Instead of asking only whether AI can complete a task, institutions should ask what completing the task used to teach, who still needs that capacity, and how the new workflow will keep it alive. Sometimes the answer will support full delegation. Sometimes it will support AI assistance plus a moment of explanation, comparison, or defense. Sometimes it will support an AI-free exercise because fluency, memory, or embodied practice is itself the point. The value of the framework is that it does not settle these choices in advance. It gives institutions a vocabulary for making the choices explicit.

This is a modest claim, but it changes the burden of justification. Once delegation is attractive, the institution that removes an occasion for practice should be able to say whether the lost practice was empty burden, replaceable support, or a necessary site where judgment was learned.

The idealization of perfect AI therefore sharpens the case for learning. Once the argument no longer depends on hallucinations, biased outputs, or fragile systems, learning appears as part of the infrastructure of accountable delegation. The goal is not to keep every person close to every technical operation. It is to preserve enough learned contact with purposes, reasons, failures, and alternatives that delegation can still be judged by those it affects. That is what learning is for when AI does the work.

\section{Ethical Considerations and Adverse Impact Statement}

The paper uses no human subjects, datasets, model training, or deployment. Its main risk is misuse: institutions might cite the argument to shift governance duties onto individuals asked to keep adapting, or to deny beneficial automation to disabled people, overburdened workers, and communities for whom AI assistance expands access. The paper's response is institutional: accountable AI-mediated work requires usable reasons, accessible contestability, and support for AI assistance when it expands agency.

\section*{LLM Usage Declaration}

During manuscript preparation, the author used large language models (LLMs) and AI agents as assistive tools for literature searching and for refining and polishing the language and structure of the text. The author has manually reviewed the manuscript and takes full responsibility for its content.

\bibliographystyle{plainnat}
\bibliography{references}
\end{document}